# Selective Hydrogen Molecule Dissociation on Ca$_2$N Monolayer

Gwan Woo Kim[1], Soonmin Jang[2], and Gunn Kim[1]

[1]*Department of Physics and HMC, Sejong University, Seoul 05006, Republic of Korea*

[2]*Department of Chemistry, Sejong University, Seoul 05006, Republic of Korea*

**Abstract**

Developing efficient hydrogen storage and conversion technologies is essential for sustainable energy. This study investigates the catalytic potential of a dicalcium nitride (Ca$_2$N) monolayer for hydrogen dissociation using density functional theory (DFT) and *ab initio* molecular dynamics (AIMD) simulations. We find that atomic hydrogen preferentially adsorbs at Ca-centered hollow sites (labeled A sites), while molecular hydrogen adsorption is limited to bridge sites (labeled B sites). Importantly, AIMD simulations reveal that H$_2$ dissociation at B sites inhibits further adsorption, suggesting a mechanism of controlled H$_2$ dissociation. The current findings emphasize the potential of pristine Ca$_2$N as a catalyst for H$_2$ dissociation-related processes and motivate future investigations of its activity in hydrogen evolution reactions.

# I. Introduction

The global search for sustainable energy solutions has positioned hydrogen molecules ($H_2$) as a promising energy carrier due to their abundance, high gravimetric energy density, and clean combustion byproducts [1-7]. However, the widespread adoption of a hydrogen-based economy depends on overcoming several critical challenges, including efficient and safe hydrogen storage, delivery, and cost-effective production [1-7]. Two-dimensional (2D) materials have emerged as highly attractive candidates due to their exceptional surface-to-volume ratios and tunable electronic structures, providing versatile platforms for hydrogen interaction [8-36].

In particular, the unique atomic-scale thickness of 2D materials provides a large surface area for efficient interactions with $H_2$ molecules, while their electronic properties can be tailored by chemical modifications or structural design. Among 2D materials, graphene - the prototypical system - has been extensively studied for hydrogen storage, mainly through physisorption [8-12]. However, the weak binding energy between $H_2$ and graphene requires enhancements such as functionalization or doping to improve hydrogen uptake [8-12]. For example, doping graphene with transition metals or introducing surface defects has been shown to significantly enhance hydrogen binding [37-40].

Besides graphene, other 2D materials such as transition metal dichalcogenides (TMDs) and MXenes have attracted increasing attention for their potential in hydrogen storage and dissociation applications. TMDs, such as $MoS_2$ and $WS_2$, provide stronger hydrogen binding compared to graphene due to the involvement of transition metal atoms [13-19]. However, their strong chemisorption of hydrogen can limit the reversibility, which poses a challenge for their application as catalysts. On the other hand, MXenes—a family of 2D transition metal carbides, nitrides, and carbonitrides—exhibit tunable electronic properties and large surface areas, allowing a balance between adsorption strength and reversibility [20-36]. Recent studies have also explored the catalytic potential of 2D materials to enhance the kinetics of hydrogen-related reactions, such as $H_2$ dissociation, further expanding their appeal for hydrogen economy applications [41-45].

Among the emerging 2D materials, 2D electrides have attracted considerable attention for their advanced electronic properties [46-49]. In particular, dicalcium nitride ($Ca_2N$) has attracted considerable interest since its discovery in 2013 [50]. With its layered structure of interstitial anionic electrons, $Ca_2N$ exhibits remarkable thermal and electrical conductivities, as well as an exceptionally low work function [51, 52]. Although $Ca_2N$ has been studied for hydrogen adsorption and dissociation [53], a comprehensive understanding of the relationship between its electronic structure and the mechanisms governing hydrogen interactions remains incomplete. Particularly, dynamic processes such as $H/H_2$ adsorption and the resulting surface modifications require further investigation to fully elucidate its potential.

This study aims to bridge this knowledge gap by investigating the hydrogen adsorption properties of $Ca_2N$ monolayers using density functional theory (DFT) calculations. Building on previous studies, we use first-principles calculations to identify optimal adsorption sites for hydrogen atoms and analyze the energetics of $H_2$ adsorption. Especially, we use ab initio molecular dynamics (AIMD) simulations to capture the electronic and structural changes associated with hydrogen adsorption and dissociation on $Ca_2N$ surfaces. In addition, we present simulated scanning tunneling microscopy (STM) topographic images to visualize surface structures before and after hydrogen interactions. Our results reveal a site-selective adsorption mechanism for hydrogen atoms on the $Ca_2N$ surface, with high-symmetry bridging sites being energetically favorable. Furthermore, AIMD simulations highlight that surface modifications induced by $H_2$ dissociation inhibit further adsorption, suggesting a potential catalytic role for $Ca_2N$ in controlling $H_2$ dissociation.

By providing a detailed analysis of the adsorption energetics, surface modifications, and catalytic potential of $Ca_2N$, this study significantly advances the understanding of its role in hydrogen evolution reactions (HERs), an essential aspect that has not been comprehensively addressed. The findings not only improve the fundamental understanding of 2D electrides in hydrogen-related applications, but also pave the way for their development as promising materials in the hydrogen economy.

## II. Computational methods and model structures

We investigated the interaction of hydrogen with the $Ca_2N$ monolayer using first-principles calculations based on DFT [45-49] as implemented in the Vienna *Ab initio* Simulation Package (VASP) [62, 63]. The projector augmented wave (PAW) method [59-61] was used to represent the core electrons, and a plane-wave basis set with a kinetic energy cutoff of 600 eV was employed. The electron exchange-correlation was described by the Perdew-Burke-Ernzerhof (PBE) generalized gradient approximation (GGA) functional [64]. The Brillouin zone was sampled using a Γ-centered Monkhorst-Pack $k$-point mesh of $17 \times 17 \times 1$ for the primitive cell and $4 \times 4 \times 1$ for the $2 \times 2 \times 1$ supercell used for adsorption studies. The Methfessel-Paxton smearing method [65] with a width of 30 meV was used for electronic structure calculations.

To investigate hydrogen adsorption, a $2 \times 2 \times 1$ supercell of $Ca_2N$ was constructed, providing a vacuum spacing of more than 14 Å between periodic images of adsorbates to minimize spurious interactions. Structural optimizations were performed until the Hellmann-Feynman forces [66] were less than 0.005 eV/Å. The Grimme-D3 van der Waals correction [67] was included to account for dispersion forces, and spin polarization was considered in all calculations. To determine the preferred adsorption sites for atomic hydrogen, we systematically placed a single hydrogen atom at various high-symmetry sites on the $Ca_2N$ surface and optimized the structures. The adsorption energy ($E_{ads}$) of atomic and molecular hydrogen was calculated as:

$$E_{ads} = E[Ca_2N + H/H_2] - E[Ca_2N] - n\, E[H],$$

where $E[Ca_2N + H/H_2]$ is the total energy of the $Ca_2N$ supercell with adsorbed hydrogen (atom or molecule), $E[Ca_2N]$ is the total energy of the pristine $Ca_2N$ supercell, and $n$ is the number of adsorbed hydrogen atoms ($n = 1$ for atomic H, and $n = 2$ for molecular $H_2$). $E[H]$ is the energy of a single hydrogen atom in its spin-polarized ground state.

The AIMD simulations were also performed using the VASP code with the Nosé-Hoover thermostat [68, 69] to investigate the dynamics of hydrogen adsorption and dissociation at 300 K. A time step of 1 fs was used, and each AIMD run consisted of 10,000 steps (10 ps). The force convergence criterion for AIMD was set to 0.01 eV/Å. The Grimme-D3 correction was included, consistent with the DFT calculations. To simulate scanning tunneling microscopy (STM) images, we used the Tersoff-Hamann approximation [70-74] as implemented in the VASP code. Careful convergence tests with respect to the plane-wave cutoff energy and $k$-point mesh were performed to ensure the accuracy of all calculations.

# III. Results and discussion

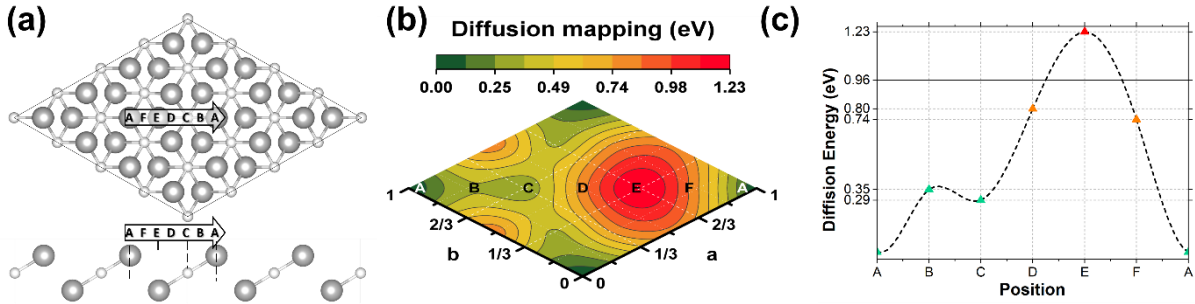

**Figure 1.** (a) Atomic model structure of the Ca$_2$N monolayer, showing key adsorption sites: hCa (A), bridge (B), hN (C), hN-ontop (D), ontop Ca (E), and hCa-ontop (F). (b) Diffusion energy map reconstructed from the adsorption energies at the six labeled sites. (c) Line profile of the diffusion energy map, with the horizontal line indicating the 0.96 eV threshold for stable adsorption.

Our DFT calculations for the Ca$_2$N primitive cell yielded a lattice constant of 3.597 Å, which is in agreement with previous theoretical studies [41-44]. The optimized structure has the following key parameters: a Ca-N bond length of 2.430 Å, a Ca-Ca z-axis distance of 2.522 Å, and a Ca-N-Ca bond angle of 95.4°. The surface structure, as shown in Figure 1(a), consists of a triangular lattice of calcium atoms defining several adsorption sites. The two primary hollow sites are the calcium-centered hollow (A), located directly above a calcium atom, and the nitrogen-centered hollow (C), located in the center of each triangular unit formed by the calcium atoms. We also considered a bridge site (B) between two surface calcium atoms and an on-top site (E) above a calcium atom. To explore potential diffusion pathways, we included two additional adsorption sites: D (between C and E) and F (between E and A), for a total of six sites studied (Figure 1a).

Our analysis focuses on the highly symmetric A, C, and B sites. The relative adsorption energies of the sites, shown in Figure 1(c), indicate that the A site is the most energetically favorable, with an adsorption energy of -0.957 eV. In contrast, the E site is highly unstable, with a positive adsorption energy of +0.268 eV. The calculated energy barrier for hydrogen transfer from the C site to the A site is remarkably low (0.057 eV), while the reverse process (A to C) has a higher energy barrier (0.347 eV). This significant difference in energy barriers strongly suggests a preferential migration of hydrogen from the C site to the more stable A site at 300 K. This preference is further supported by our AIMD simulation results presented later. The steep slope found in the energy profile around the D site suggests a low probability of hydrogen atoms residing in this region, indicating that this region is unlikely to be a stable adsorption site or a significant intermediate state in diffusion pathways.

Our study of molecular hydrogen ($H_2$) adsorption, including in-plane relaxation, reveals the dissociation mechanism on $Ca_2N$. We found that dissociation occurs when $H_2$ approaches the surface with its bond axis nearly parallel to the $Ca_2N$ plane and its center of mass directly above the B site, initiating at a vertical distance of 1.671 Å from the surface Ca atom. After dissociation at the B site, the hydrogen atoms migrate to the more stable A and C sites, forming a co-adsorption state with a combined binding energy of -1.471 eV. This binding energy is significantly larger (more stable) than the sum of the individual atomic adsorption energies (-0.957 eV at the A site and -0.668 eV at the C site), indicating an additional stabilization of 0.077 eV per hydrogen atom upon co-adsorption. This enhanced stability is attributed to surface relaxation and rearrangement upon co-adsorption, highlighting the importance of surface dynamics in the stabilization process. This relaxation may play a critical role in facilitating HER. To further elucidate the molecule-surface interactions, we also investigated $H_2$ adsorption at the highly symmetric A and C sites. While adsorption of $H_2$ at the B site consistently led to dissociation and occupation of the A and C sites, direct adsorption of $H_2$ at the A site, the most stable site for atomic hydrogen, yielded a distinct result, which will be discussed in the following section.

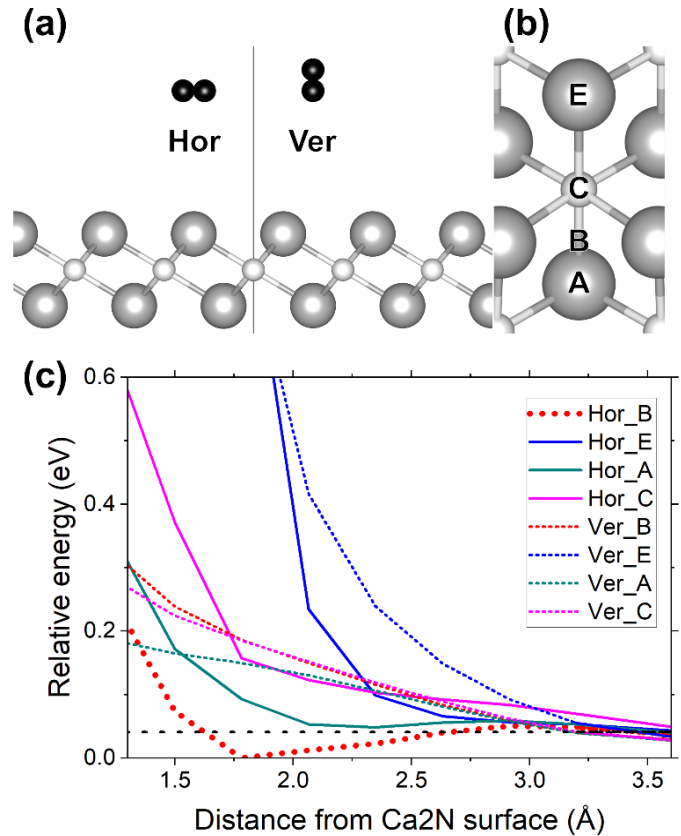

**Figure 2.** (a) Side view of the $H_2$ molecule in both horizontal and vertical orientations relative to the $Ca_2N$ surface, illustrating the various adsorption configurations. (b) Top view of the four key configurations identified during the adsorption simulations. (c) Adsorption energy landscape to determine the most favorable $H_2$ adsorption structure on the $Ca_2N$ surface, obtained by investigating all eight high-symmetry initial configurations on the $Ca_2N$ surface.

We investigated various orientations and adsorption pathways of $H_2$ molecules on the $Ca_2N$ surface. As shown in Figure 2(a), we considered configurations with the molecular axis oriented either parallel (horizontal) or perpendicular (vertical) to the surface plane. Initially, we hypothesized that horizontally oriented molecules might initially show physisorption-like behavior at non-bridge sites and subsequently diffuse to a more energetically favorable bridge configuration between two Ca atoms. This hypothesis was based on the possibility of a weak initial interaction at various sites before reaching the preferred bridging configuration. However, our simulations revealed a significant repulsive interaction between the $H_2$ molecule and the $Ca_2N$ surface at interatomic distances below ~3.3 Å. This repulsive interaction, overcoming the van der Waals attraction, resulted in molecular tilting and stabilization at a distance of ~3.7 Å from the surface. This indicates the presence of a repulsive potential energy barrier inhibiting direct chemisorption over most of the $Ca_2N$ surface, with the B site being a notable exception. We identified a distinct pathway for $H_2$ dissociative adsorption: the molecule approaches the B site, undergoes bond scission (dissociates), and the resulting H atoms chemisorb at the A and C sites, consistent with our previous findings regarding the preferred adsorption sites for atomic hydrogen. This confirms the B site as a key intermediate state in the $H_2$ dissociation process.

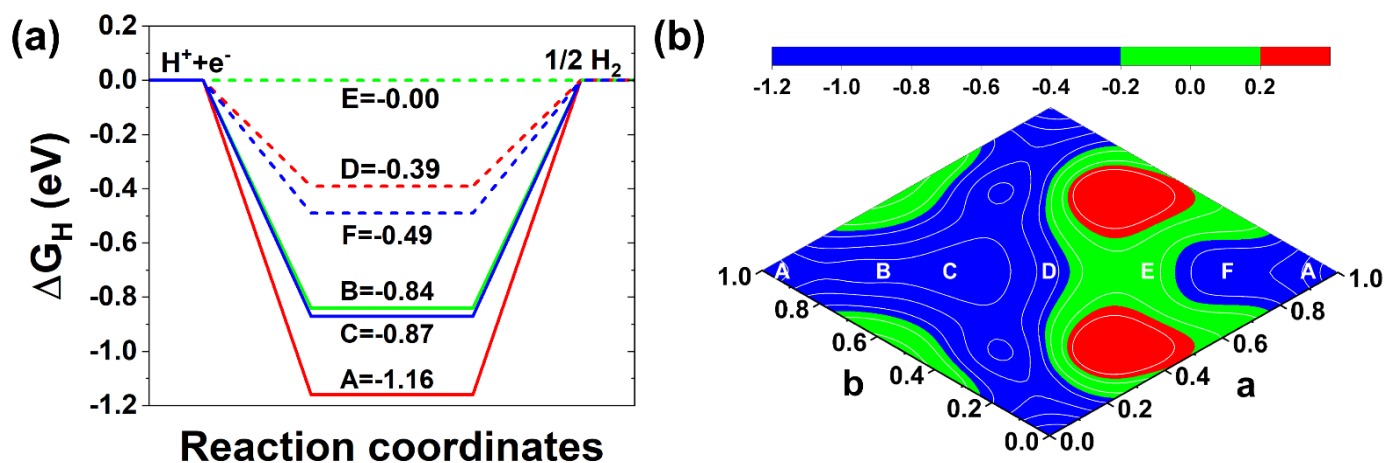

**Figure 3.** Gibbs free energy analysis for HER on $Ca_2N$. (a) Calculated Gibbs free energy at specific adsorption sites. (b) Estimated Gibbs free energy distribution across the $Ca_2N$ surface, showing that approximately 28% of the surface area (green) exhibits favorable values for HER.

We investigated the feasibility of $Ca_2N$ as a catalyst for HER by analyzing the Gibbs free energy at six different surface positions (A to F), a key parameter influencing HER activity. As shown in Figure 3, point E exhibits the Gibbs free energy closest to zero, despite being the most unstable in terms of adsorption energy. This is consistent with the well-established notion that overly strong binding sites can be detrimental to HER. However, when the analysis was extended to the entire surface, it was found that 28% of the area could have a favorable Gibbs free energy for HER (Figure 3). Considering the adsorption energy previously analyzed (figure not shown, but referenced earlier), the actual effective area for HER is likely much smaller, possibly a single point. This suggests limitations in the use of monolayer $Ca_2N$ as an efficient HER catalyst.

To further understand the initial approach of $H_2$ molecules prior to dissociation and preferred adsorption sites, we performed additional calculations. The hydrogen molecules were placed at different positions (A-F) in both vertical and horizontal orientations, constraining H atoms to prevent dissociation. Notably, only horizontally oriented molecules approaching the B site ($H_2$@B) showed a significant stability point near the energy barrier (~30 meV) in Figure 2(b). While $H_2$@A appears to be more stable in a certain region, this is less important due to the unfavorable (endothermic) nature of H$_2$@A adsorption. None of the remaining six pathways showed well-defined stable structures.

To understand the mobility of adsorbed hydrogen molecules, we analyzed the adsorption energies of eight high-symmetry configurations. This analysis revealed a strong energetic preference for horizontal molecular orientation on the Ca$_2$N surface, especially at the B site, suggesting a mechanism involving lateral diffusion. Conversely, significant energy barriers hinder vertical adsorption. Our AIMD simulations confirm this mechanism, with the final positions of the dissociated hydrogen atoms agreeing with the calculated adsorption energies (see Supplementary Material).

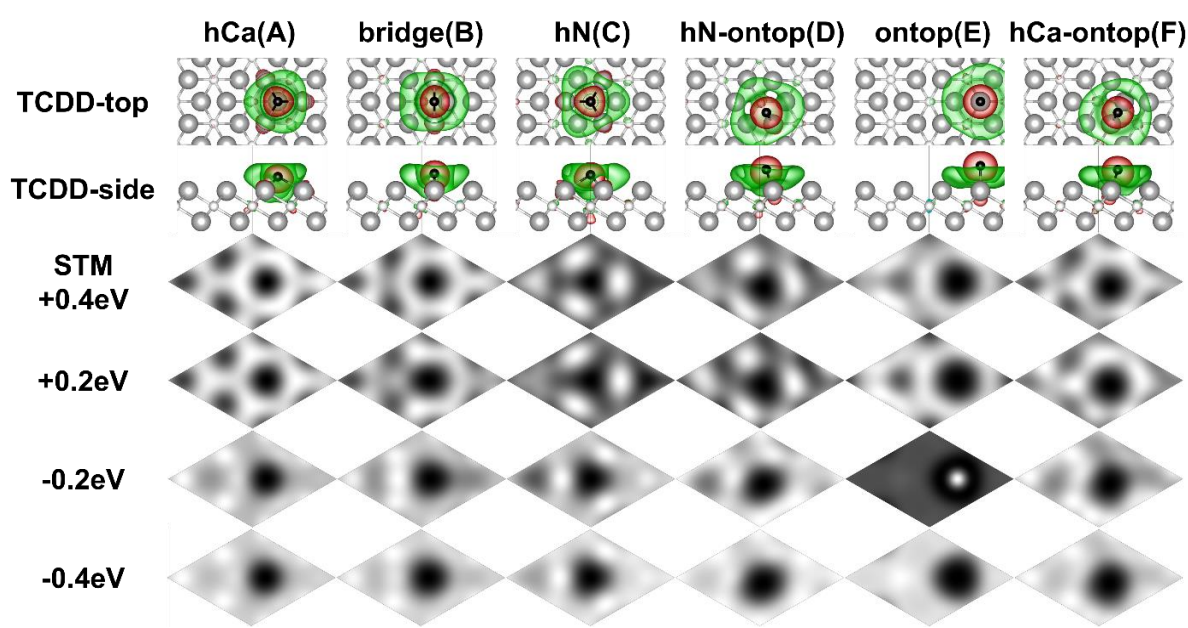

**Figure 4.** Total charge density difference (TCDD) plots and corresponding scanning tunneling microscopy images showing hydrogen atom adsorption on the Ca$_2$N surface at various positions. The energy values indicated on the left represent the relative energy difference from the Fermi level. Our STM simulations mimic the constant height mode, where the height is assumed to be 2.8 angstroms along the z-axis between the tip and the closest hydrogen atom.

To understand the electronic structure changes upon hydrogen adsorption, we analyzed total charge density difference (TCDD) plots and simulated constant-height STM images [61] for the atomic and molecular adsorption configurations shown in Figure 4. The STM images were normalized, with white and black representing the highest and lowest tunneling currents, respectively. For atomic hydrogen adsorption, distinct electronic signatures are observed at the most stable adsorption sites, A and C. This is evident in the simulated STM images at +0.4 eV bias, where the spatial distribution of the tunneling current differs significantly between these two configurations. Similar distinctions are observed between sites C and D (and also between A and F). The STM image for the B site is similar to that for A, but exhibits an asymmetry in the bright features, forming a pseudo-hexagonal pattern. This indicates that, although experimentally challenging, distinguishing between sites A and C via STM is feasible. The unstable on-top site (E) exhibits a markedly different electronic structure, characterized by a prominent central hexagonal feature in the simulated STM image. At a bias of -0.4 eV, contributions from the adsorbed hydrogen atoms and surface anionic electrons become faintly visible, although these features are less pronounced than at +0.4 eV. This reduced contrast at negative bias is consistent with previous STM studies on similar systems [62]. Nevertheless, even at -0.4 eV, sites A and C can be distinguished by the shape of the dark regions in the STM images: approximately circular for A and triangular for C. These simulated STM features provide valuable guidance for future experimental verification of our computational predictions, facilitating the identification of specific adsorption sites on hydrogen-adsorbed $Ca_2N$ surfaces.

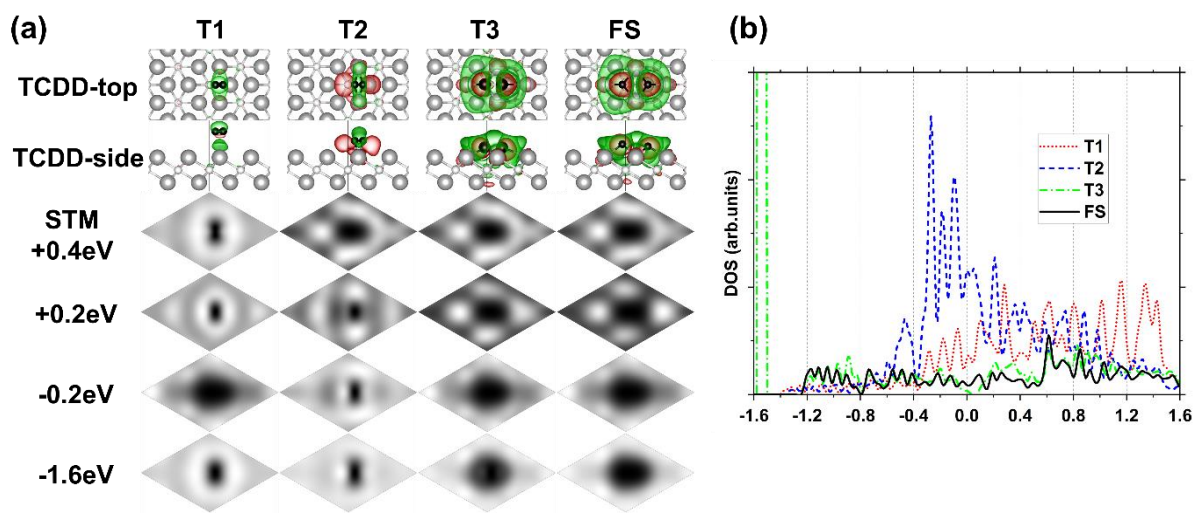

**Figure 5.** (a) Total charge density difference (TCDD) plots and corresponding simulated STM images for the T1, T2, T3, and FS configurations. Energies relative to the Fermi level are shown on the left. (b) Density of states (DOS) plots for the molecular adsorbates. Only the T3 configuration exhibits a state within the band gap.

Simulated STM images were analyzed consistently for all four molecular adsorption processes to investigate the adsorption mechanism of hydrogen molecules on $Ca_2N$ surfaces. The final atomic structures observed in the -1.6 eV STM images of the T3 configuration are similar to those of atomic hydrogen adsorption (the influence of the dissociated hydrogen molecules is visible as a weak contrast change in the simulated STM images). This suggests a cooperative adsorption mechanism: pre-adsorbed hydrogen atoms, by bonding with neighboring Ca atoms, induce local structural changes that influence subsequent adsorption events. This mechanism is in agreement with our adsorption energy calculations. The metallic character of $Ca_2N$, with its significantly higher electron density compared to hydrogen atoms or molecules, presents a challenge for the experimental resolution of electronic structure changes near the Fermi level upon hydrogen adsorption. Nevertheless, a distinct density of states (DOS) peak associated with the hydrogen molecule is observed within the $Ca_2N$ bandgap (around -1.5 eV) during the intermediate stages of adsorption (prior to complete structural relaxation). Direct experimental observation of these molecular adsorption processes is challenging due to their transient nature. While techniques such as STM and spectroscopy [63-65] have limitations in capturing these transient states, atomic force microscopy (AFM) may offer some potential for this purpose, although the resulting structures are predicted to be unstable. Based on our findings, we investigated the possibility of subsequent adsorption of hydrogen molecules following the initial dissociation and adsorption of H atoms at the A and C sites adjacent to the B site. In particular, we determined whether another $H_2$ molecule could adsorb to a neighboring B site, potentially interacting with the pre-adsorbed hydrogen atoms to form new $H_2$ molecules. To address this issue, we performed *ab initio* molecular dynamics (AIMD) simulations.

Our AIMD simulations indicate that subsequent $H_2$ adsorption at the same B site is energetically unfavorable. Upon dissociation and adsorption of hydrogen atoms at the neighboring A and C sites of the B site, the $Ca_2N$ lattice undergoes significant local deformation. The Ca-Ca distance between neighboring Ca atoms contracts from 3.569 Å to 3.482 Å. The local lattice constant around the B site experiences an even more pronounced contraction, decreasing to 3.249 Å due to the presence of the adsorbed hydrogen atoms. This local contraction alters the Ca-Ca inter-bridge spacing, which significantly modifies the interaction potential for subsequent $H_2$ adsorption. This contraction is attributed to the formation of covalent bonds between the hydrogen atoms and the surrounding Ca atoms, which changes the local electronic structure and induces a net attractive interaction between the hydrogen atoms and the $Ca_2N$ lattice. Consequently, the energy required for further lattice deformation to accommodate another $H_2$ molecule at the same B site becomes prohibitively high.

Our AIMD simulations show that a horizontally oriented $H_2$ molecule approaching a B site occupied by pre-adsorbed hydrogen atoms experiences a repulsive interaction. This repulsion directs the incoming $H_2$ molecule away from the initially occupied B site to a neighboring B site. Consequently, the dissociated hydrogen atoms from this subsequently adsorbed molecule are predicted to preferentially occupy the A sites rather than the C sites. This repulsive interaction results from the electron density overlap between the pre-adsorbed hydrogen atoms and the approaching $H_2$ molecule. This electron density overlap results in an increase in the energy of the system, making direct approach to the already occupied B site energetically unfavorable. A neighboring, unoccupied B site with reduced electron density overlap thus provides a more favorable adsorption site. Also, the current AIMD simulations at room temperature (300 K) and high hydrogen molecule concentration reveal the dynamic nature of atomic hydrogen adsorption on $Ca_2N$. Thermal vibrations cause the initially adsorbed hydrogen atoms at the C site to readily move to the more stable A site. Initially, these hydrogen atoms occupy the C site, but after some time (500 fs at 300 K), all the initially adsorbed hydrogen atoms migrate and become stable on A, which draws attention to the important role of thermal vibration in atomic hydrogen adsorption on $Ca_2N$. The C site provides a slightly higher energy state than the A site due to the difference in the local electronic environment. At room temperature, thermal vibrations provide enough energy for the hydrogen atom to overcome this energy barrier and move to the more stable A site. This dynamic behavior reflects the interplay between the local energy landscape and thermal vibrations.

Initially, hydrogen atoms occupy the C site, a metastable state relative to the A site. However, within 500 fs at 300 K, we observed a complete migration of all initially adsorbed hydrogen atoms to the energetically more favorable A sites. This rapid transition emphasizes the important role of thermal activation in the adsorption kinetics. The energy difference between the C and A sites results from variations in their local electronic environments. At room temperature, the available thermal energy ($k_BT \approx 25$ meV) is sufficient to overcome the relatively small energy barrier separating these sites, promoting the observed migration. This dynamic behavior reveals the interplay between the potential energy landscape, defined by the electronic structure, and thermal fluctuations, which govern the adsorption dynamics.

Analysis of the AIMD trajectories (every 10,000 frames) reveals a consistent dissociation mechanism: $H_2$ molecules approach the B site with a preferred orientation, forming an angle of about 120° with the surface. At the B site, $H_2$ dissociates, and the resulting H atoms migrate to the A and C sites. The H atoms migrating to the A sites rapidly adsorb to the surface, while those at the C sites recombine to form $H_2$ and then desorb. This consistent behavior, observed in simulations at 300 K, strongly supports our DFT results and predictions and demonstrates the robustness of the underlying $H_2$ dissociation and adsorption mechanism on $Ca_2N$ surfaces across different temperature regimes.

The preferred approach orientation and the ~120° angle suggest a specific interaction between $H_2$ and the $Ca_2N$ surface, reflecting a balance between repulsive and attractive forces. The rapid adsorption at the A sites indicates a strong interaction and a favorable energetic state, likely due to covalent bond formation with the $Ca_2N$ surface. On the other hand, the recombination and desorption of H atoms at the C sites indicate a less favorable energetic state for atomic adsorption compared to the molecular form, possibly due to weaker or fewer bonding interactions with the surface. This recombination and desorption process appears to be less sensitive to temperature variations, suggesting that it is primarily determined by the local electronic structure and bonding environment.

## IV. Conclusion

In an attempt to understand the hydrogen molecule adsorption and dissociation deeply on the 2-dimensional $Ca_2N$ surface, we have performed a series of DFT calculations and AIMD simulations. In summary, our combined DFT and AIMD investigations elucidate the mechanism of hydrogen interaction with the $Ca_2N$ surface. We find that $H_2$ adsorption is highly selective to B sites and that subsequent adsorption is hindered by local lattice distortions induced by pre-adsorbed hydrogen. While this limits the overall adsorption capacity, the selective adsorption and dissociation mechanism holds promise for catalytic applications in hydrogen storage and fuel cell technologies. The abundance and low cost of Ca and N compared to the platinum group metals (Pt, Pd, Ir, and Ru) further enhance the attractiveness of $Ca_2N$ as a potential catalyst. [66] Preliminary HER calculations suggest promising activity, but further theoretical and experimental studies are needed to fully evaluate its catalytic performance. This work provides a fundamental understanding of the hydrogen-$Ca_2N$ interaction and establishes a foundation for future research exploring $Ca_2N$ as a potential material in hydrogen energy technologies, including as storage and/or catalyst for HER.

## Acknowledgments

J.C. and G.K. acknowledge the Basic Science Research Program's financial support (Grants No. NRF-2019R1F1A1058177 and No. NRF-2020R1A6A1A03043435) through the National Research Foundation of Korea (NRF) funded by the Government of Korea.